\documentclass{article}
\usepackage{authblk}
\usepackage{graphicx}
\usepackage{amsmath}
\usepackage{amssymb}
\usepackage{physics}
\usepackage{tabularx}
\usepackage[export]{adjustbox}
\usepackage{cite}

\title{Photonic Chern insulators from two-dimensional atomic lattices interacting with a single surface plasmon polariton}
\author[1]{Rituraj}
\author[2]{Meir Orenstein}
\author[3]{Shanhui Fan}
\affil[1,3]{Department of Electrical Engineering, Stanford University, Stanford, California 94305, USA}
\affil[2]{Department of Electrical Engineering, Technion-Israel Institute of Technology, Haifa 32000, Israel}
\affil[3]{shanhui@stanford.edu}

\date{}
\begin{document}
\maketitle
\section*{Abstract}
We study the polaritonic bandstructure of two-dimensional atomic lattices coupled to a single excitation of a surface plasmon polariton mode. We show the possibility of realizing topological gaps with different Chern numbers by having resonant atomic transitions to excited states with different angular momentum. We employ a computational method based on the recently proposed Dirichlet-to-Neumann (DtN) map technique which accurately models non-Markovian dynamics as well as interactions involving higher-order electric and magnetic multipole transitions. We design topologically robust edge states which are used to achieve unidirectional emission and non-reciprocal transmission of single photons. We also point out the challenges in realizing bands with higher Chern numbers in such systems.

\section{Introduction}
Periodic lattice of atoms interacting with a nanophotonic environment has gained increasing interests in recent years with the advances in the related quantum technology \cite{masson2020atomic, corzo2019waveguide, perczel2017topological, bettles2017topological, perczel2017photonic, asenjo2017exponential, bettles2016enhanced, facchinetti2016storing, shahmoon2017cooperative, rituraj3}. Such a periodic system has been shown to exhibit several interesting phenomena arising from coherent collective oscillations such as sub- and super-radiance, near perfect reflection of radiation, forbidden energy gaps \cite{facchinetti2016storing, asenjo2017exponential, bettles2016enhanced, shahmoon2017cooperative, shen2005coherent, rituraj3} etc. A one-dimensional (1D) lattice of atoms can act as a waveguide for single photons and a quantum emitter placed near the atomic lattice can emit selectively into the guided modes \cite{masson2020atomic}. Similarly, a 2D atomic lattice supports surface confined Bloch modes in the single excitation regime \cite{perczel2017photonic, asenjo2017exponential}. In recent works we illustrated the design of atom based mirrors, cavities and isolated flat bands for manipulating single photons in 2D surface plasmon polariton (SPP) mode \cite{rituraj2, rituraj3}. There have also been a few recent theoretical works proposing the design of 2D atomic lattices having polaritonic bands with non-trivial topology allowing for topologically robust edge states \cite{perczel2017topological, bettles2017topological, perczel2017photonic}. All these results point to the exciting possibility of realizing fully atom based optical interfaces. 

However, most of these works simplify the atom-photon interaction Hamiltonian by tracing out the photon degrees of freedom and applying the Markov approximation \cite{perczel2017photonic,asenjo2017exponential}. This approximation is not valid when the electromagnetic environment has narrow bandwidth features and is highly dispersive. The Markov approximation is also violated when the length of a spontaneously emitted photon becomes comparable to or smaller than the atom spacing, which might occur for stronger atom-photon coupling resulting in faster decay rates. Another almost universal approximation in these models is the electric dipole approximation based on the assumption that the photon wavelength is much longer than the atomic size \cite{scully1999quantum, walls2007quantum}. This approximation may not hold for surface plasmon polaritons with wavelengths far smaller than the free-space wavelengths.

In this work we study the interaction of a 2D atomic lattice with a SPP mode focusing on the emergent topological properties in the single excitation regime using a recently proposed model which does not suffer from the limitations of the aforementioned Markov and dipole approximations \cite{rituraj3}. Strongly confined SPP mode provides a significantly enhanced atom-photon interaction strength, especially near the surface plasmon frequency where the photonic density of states diverges \cite{rivera2016shrinking}. The SPP mode is highly dispersive and has a very sharp cutoff near the surface plasmon frequency. Also, near the surface plasmon frequency the plasmon wavelength can become comparable to the atomic size, more-so for the case of Rydberg atoms in highly excited states, or large quantum dots. Thus, both the Markov and dipole approximations are not valid when the atomic transition frequency lies close to the surface plasmon frequency, signifying the usefulness of our general model. We further extend this previously proposed model to compute topological properties of the compound system. We compute the polaritonic band dispersion and the associated Chern number for an infinite atomic lattice and also show the existence of topologically robust unidirectional edge states at the interface between two lattices with different Chern numbers. Using these edge states, we show the possibility of achieving non-reciprocal transmission and unidirectional emission from single photon sources.

The rest of the paper is organized as follows. In Section 2, we describe the atom-SPP system and provide an overview of the Dirichlet-to-Neumann (DtN) map based bandstructure computational method. In Section 3, we design and investigate three different atomic lattices which have topological gaps with Chern number 0, 1 and 2 respectively. Subsequently, using the atomic lattices with Chern numbers $0$ and $1$, we realize topologically robust unidirectional edge states. Finally, after a brief discussion of the challenges involved in realizing bands with higher Chern numbers, we conclude in Section 4.

\section{Mathematical Formulation}
\begin{figure}
\centering
\includegraphics[width = 0.8\linewidth]{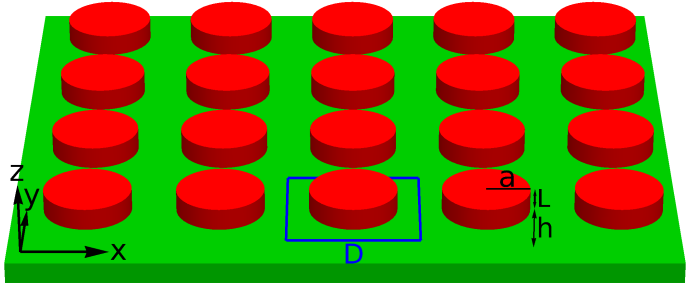}
\caption{A square lattice of atoms (red cylinders) with periodicity $D$ placed at a distance $h$ from the surface supporting a SPP mode (shown in green).}
\end{figure}

\subsection{SPP and Atom model}
A periodic atomic lattice coupled to a SPP mode is shown schematically in Fig.\,1. In this work, we consider a square lattice but the formalism is general and applicable to any lattice type. The infinite 2D surface supporting the SPP mode is taken to be the $z=0$ plane (shown in green). The atoms in the lattice (represented by red cylinders) are indexed by $(u,v)$ and are placed at coordinates $(\textbf{r}_{uv}, h)$ where $\textbf{r}_{uv} = (uD, vD)$ is the in-plane coordinate, $D$ is the lattice period and $h$ is the atom-surface distance. We assume that the atoms are separated from the surface by vacuum, and the SPP mode is ideal without propagation losses. As we will see later, a few lattice periods are enough to exhibit the topological phenomena such as unidirectional edge states arising from the periodic structure. Thus the lossless SPP model is a good approximation as long as the SPP propagation length is larger than few lattice periods. In the Coulomb gauge, the SPP vector potential operator $\textbf{A}(\textbf{r},z)$ in the upper half space ($z > 0$) is given by \cite{tame2013quantum, archambault2010quantum, ferreira2020quantization, blow1990continuum}:
\begin{equation}
\textbf{A}(\textbf{r},z) = \iint d^2\textbf{k}\underbrace{\frac{1}{2\pi}\sqrt{\frac{\hbar}{2L_\textbf{k}\epsilon_0 \omega_\textbf{k}}} \left ( i\hat{\textbf{k}} - \frac{k}{\kappa}\hat{\textbf{z}} \right )e^{-\kappa z}e^{i\textbf{k}\cdot\textbf{r}}}_{\begin{matrix}\textbf{A}_\textbf{k}(\textbf{r}, z)\end{matrix}}  a_\textbf{k} + H.c, 
\end{equation}
where, $a_\textbf{k}, a_\textbf{k}^{\dagger}$ are the bosonic annihilation and creation operators for the SPP mode photon satisfying the commutation $[a_\textbf{k},\   a^{\dagger}_{\textbf{k}'}] = \delta^2 (\textbf{k-k}') $, $\hbar$ is the reduced Planck's constant, $\epsilon_0$ is the vacuum permittivity, $\kappa = \sqrt{k^2 - \omega_\textbf{k}^2/c^2}$ is the spatial decay rate of the mode along $z$, $L_\textbf{k}$ is the characteristic modal dimension given by $L_\textbf{k} = (\kappa ^2 + k^2)/\kappa ^3$ and is derived by normalization consideration, H.c stands for Hermitian conjugate. $\omega_\textbf{k}$ is the SPP mode (angular) frequency for in-plane wavevector $\textbf{k} \equiv (k_x, k_y)$, and could be well approximated for frequency close to the surface plasmon frequency $\omega_{sp}$ by:
\begin{equation}
\omega_\textbf{k} = \sqrt{\frac{1 + \epsilon}{\epsilon}}ck \approx \frac{\omega_p}{\sqrt{2}} (1 - \frac{k_p^2}{8k^2}) = \omega_{sp} - \frac{\beta}{k^2} ,
\end{equation}
where, $\epsilon$ is the real dielectric constant of the metal given by the Drude model, $\omega_p$ is its plasma frequency, $k_p$ is the free space wave-vector magnitude at frequency $\omega_p (= ck_p)$, $c$ is the speed of light in vacuum \cite{economou1969surface, jablan2009plasmonics, jablan2013plasmons}.

For simplicity, we model the atom as an infinite potential well confining the electron in a cylinder of radius $a$ and height $L$ as shown in Fig.\,1. In this paper, we interchangeably use atoms and quantum dots to refer to this quantum system with discrete energy levels. The electron wavefunction for state $(l,m,n)$ in the cylindrical quantum dot is given by:
\begin{equation}
\psi_{l,m,n}(\textbf{r},z) = \sqrt{\frac{2}{L}}\sin{\frac{l(z-h)}{L}}\frac{1}{\sqrt{\pi}aJ_{m+1}(j_{mn}a)}J_m(j_{mn}r/a)e^{im\theta}\Theta(r,z),
\end{equation}
where, $l$ is a positive integer, $J_m$ is the Bessel function of the first kind of order $m$, $j_{mn}$ is its $n^{th}$ zero, $\Theta(r,z)$ is a scalar function which is unity inside the cylinder $(r < a, h < z < h + L)$ and zero outside \cite{baltenkov2016electronic}, and $\theta$ is the azimuthal angle. Note that the state $(l, m, n)$ is an eigenstate of the $z$ angular momentum operator $(L_z)$ with eigenvalue $m\hbar$. In the presence of a static magnetic field $B_0\hat{\textbf{z}}$, the degeneracy between the $+m$ and $-m$ states is broken (Zeeman splitting) and the energy eigenvalue is given by $E_{l,m,n} = l^2h^2/(8m_eL^2) + \hbar^2 j_{mn}^2/(2m_e a^2) + \mu_BmB_0$, where $m_e$ is the electron rest mass and $\mu_B$ is the Bohr magneton. Since, later we will adopt a minimal coupling Hamiltonian to describe the atom-photon interaction and ignore the spin-flip processes, the electron spin contribution in the Zeeman splitting here is inconsequential as it does not affect the energy difference between the states with identical spin. We choose the quantum dot dimensions, such that the transition energy between the ground state $\ket{g}\equiv(1, 0, 1)$ and excited states $\ket{\pm}\equiv(2, \pm |m_0|, 1)$ lies close to the surface plasmon energy $\hbar\omega_{sp}$ for a particular $m_0$. We approximate the atom as a three (two) level system for $m_0\neq 0$ ($m_0 = 0$) and retain only the eigenstates resonant with the surface plasmon energy. As we will see later, in the presence of a large magnetic field (a few Tesla), even the three-level atom essentially acts as a two-level system in the frequency range of interest and the topological properties of the periodic atomic lattice depend only on the resonant $m_0$ state. 

\subsection{Minimal coupling Hamiltonian}
We use the standard minimal coupling Hamiltonian for atom-photon interaction as given by \cite{rituraj1}:
\begin{equation}
\begin{split}
H &= \underbrace{\iint d^2\textbf{k}\,\hbar\omega_{\textbf{k}}a_{\textbf{k}}^\dagger a_{\textbf{k}}}_{\begin{matrix}H_{SPP}\end{matrix}} + \sum_{u,v}\underbrace{\left( E_g f_{g u v}^{\dagger} f_{guv} + E_+ f_{+uv}^{\dagger} f_{+uv} +E_- f_{-uv}^{\dagger} f_{-uv}\right)}_{\begin{matrix}H_{atom(u,v)}\end{matrix}} \\
&+ \sum_{u,v}\underbrace{\iint d^2\textbf{k} \left(V_{+uv}(\textbf{k}) a_{\textbf{k}}^{\dagger} f_{guv}^{\dagger}f_{+uv} + V_{-uv}(\textbf{k}) a_{\textbf{k}}^{\dagger} f_{guv}^{\dagger}f_{-uv} +H.c \right)}_{\begin{matrix}H_{int(u,v)}\end{matrix}}, 
\end{split}
\end{equation}  
where, $E_g$ and $E_\pm$ are the ground and excited states' energies, and $f_{guv}$, $f_{guv}^{\dagger}$, $f_{\pm uv}$, $f_{\pm uv}^{\dagger}$ are the respective fermionic annihilation and creation operators for the electron in the atom $(u, v)$. $V_{\pm uv}(\textbf{k})$ is the atom-photon coupling strength for SPP wavevector $\textbf{k}$ given by:
\begin{equation}
V_{\pm uv}^*(\textbf{k}) = -\frac{e}{m_e}\bra{\pm_{uv}}\textbf{A}_{\textbf{k}}\cdot\textbf{p}_{uv}\ket{g_{uv}},
\end{equation}
where, $e$ is the electron charge and $\textbf{p}_{uv}=-i\hbar\vec{\nabla}_{uv}$ is the canonical momentum operator for the electron in the atom $(u, v)$. We compute $V_{\pm uv}(\textbf{k})$ without making the electric dipole approximation and thus it includes contributions from electric and magnetic multipoles of all possible orders. In the above Hamiltonian, we have ignored the terms related to intrinsic spin angular momentum and the $\textbf{A}^2$ term. We have also made the usual rotating wave approximation in the interaction Hamiltonian \cite{scully1999quantum, walls2007quantum, power1978nature}. These approximations are justified in the weak coupling regime (small $V_\textbf{k}$), which is the case here as we have a single photon interacting with a low density atomic lattice such that the period is significantly larger than the size of the atoms \cite{rituraj1}. In the two-level limit obtained by ignoring the off-resonant excited state ($\ket{+}$ or $\ket{-}$, depending upon the direction of applied static magnetic field), the Hamiltonian in Eq.\,(4) can be shown to be similar to the Hamiltonian considered in \cite{karzig2015topological} for a coupled exciton-photon system confined to two dimensions. But as shown later, opening up a complete band gap is relatively straightforward with the atomic lattice as opposed to the exciton based system. From Eqs.\,(1,3 \& 5), $V_{\pm uv}(\textbf{k})$ can be seen to be of the form $g_{\pm uv}(k)e^{\pm im_0\theta_\textbf{k}}$ ($\textbf{k} \equiv (k, \theta_\textbf{k})$ in polar coordinates) and thus has a winding which, as we will see later, leads to a non-trivial topology depending on the value of $m_0$. Note that, since the atoms do not directly (electronically) interact with each other, the topological properties of the bands of the atomic lattice can be attributed to be arising solely from the atom-SPP coupling.

To compute the eigenstates of the Hamiltonian in Eq.\,(4), we transform it into spatial representation by defining spatial bosonic creation and annihilation operators by:
\begin{equation}
c(\textbf{r}) = \frac{1}{2\pi}\iint d^2\textbf{k}\, e^{i\textbf{k}\cdot\textbf{r}} a_\textbf{k} \ ; \ c^{\dagger}(\textbf{r}) = \frac{1}{2\pi}\iint d^2\textbf{k}\, e^{-i\textbf{k}\cdot\textbf{r}} a^{\dagger}_\textbf{k},
\end{equation}
and using the SPP dispersion approximation of Eq.\,(2). The Hamiltonian then becomes:
\begin{equation}
\begin{split}
H &= \hbar\omega_{sp} \iint d^2\textbf{r} c^{\dagger}(\textbf{r}) c(\textbf{r}) + \hbar \beta \iint d^2\textbf{r}\iint d^2\textbf{r}' \frac{\ln{\lvert \textbf{r}-\textbf{r}'\rvert}}{2\pi} c^{\dagger}(\textbf{r})c(\textbf{r}') + \sum_{u,v}H_{atom(u,v)}\\
&+ \sum_{u,v}\iint d^2\textbf{r} \left(V_{+uv}(\textbf{r}) c^{\dagger}(\textbf{r}) f_{guv}^{\dagger}f_{+uv} + V_{-uv}(\textbf{r}) c^{\dagger}(\textbf{r}) f_{guv}^{\dagger}f_{-uv} +H.c \right),
\end{split}
\end{equation}
where, $V_{\pm uv}(\textbf{r})$ is the Fourier transform of the coupling strength $V_{\pm uv}(\textbf{k})$: 
\begin{equation}
V_{\pm uv}(\textbf{r}) = \frac{1}{2\pi}\iint d^2\textbf{k}\, e^{i\textbf{k}\cdot\textbf{r}} V_{\pm uv}(\textbf{k}).
\end{equation}
$V_{\pm uv}(\textbf{r})$ has the same winding in real space as $V_{\pm uv}(\textbf{k})$ does in $\textbf{k}$ space. Since all the atoms are identical, from Eqs.\,(1) and (5), $V_{\pm uv}(\textbf{k}) = e^{-i\textbf{k}\cdot\textbf{r}_{uv}}V_{\pm 00}(\textbf{k})$. Thus, we have $V_{\pm uv}(\textbf{r}) = V_{\pm uv}(\textbf{r} - \textbf{r}_{uv})$, which is the periodicity condition. Now, we solve for the Bloch eigenstates of the Hamiltonian in Eq.\,(7) in the single excitation regime, which is of the form:
\begin{equation}
\ket{\psi_\textbf{q}} = \left(\iint d^2\textbf{r}\,\phi_\textbf{q}(\textbf{r})c^{\dagger}(\textbf{r}) + \sum_{u,v} e_{+uv}f_{+uv}^{\dagger}f_{guv} + e_{-uv}f_{-uv}^{\dagger}f_{guv}\right )\ket{g,\,g,\ldots,0}
\end{equation}
where, $\textbf{q}$ is the Bloch wavevector and $\ket{g,\,g,\ldots,0}$ is the state with all the atoms in the ground state and zero photon in the SPP mode. Eq.\,(9) represents a complete basis for the system \cite{shen2005coherent, yudson2008multiphoton}. From the eigenvalue equation $H\ket{\psi_\textbf{q}} = E\ket{\psi_\textbf{q}}$, we obtain the following equations for the photon field $\phi_\textbf{q}(\textbf{r})$ and the excited state amplitudes $e_{\pm uv}$:
\begin{equation}
\hat{\Delta} \phi_\textbf{q}(\textbf{r}) + k^2 \phi_\textbf{q}(\textbf{r}) = -\frac{k^2}{\hbar\beta}\sum_{u,v} \left( e_{+uv} \hat{\Delta} V_{+uv}(\textbf{r}) + e_{-uv} \hat{\Delta} V_{-uv}(\textbf{r})\right )
\end{equation}
\begin{equation}
\hbar\left(\omega - \Omega_\pm \right )e_{\pm uv} = \iint d^2\textbf{r}\,V_{\pm uv}^*(\textbf{r})\phi_\textbf{q}(\textbf{r}),
\end{equation}
Here, we have substituted $E = \hbar\omega + \sum_{u,v}E_g$. $\hat{\Delta}\equiv \partial^2/\partial x^2 + \partial^2/\partial y^2$ is the 2D Laplacian operator, $\Omega_\pm = (E_\pm - E_g)/\hbar$ are the atomic transition frequencies corresponding to $\ket{\pm}$ states, and $\omega$ and $k$ are the SPP frequency and wavevector magnitude respectively related by the dispersion relation in Eq.\,(2). Note that, the SPP wavevector magnitude $k$ is different from the Bloch wavevector magnitude $q$. Since the atom-field interaction is local, $V_{\pm uv}(\textbf{r}) = 0$ for $|\textbf{r} - \textbf{r}_{uv}| > D/2$. Thus, one can solve Eqs.\,(10,11) inside a unit cell, say the one containing atom $(0, 0)$, to get a complete and orthogonal basis set of photon field modes $\phi_{k,m}(\textbf{r})$:
\begin{equation}
\phi_{k,m}(\textbf{r}) = J_m(kr)e^{im\theta} + b_mH_m^{(1)}(kr)e^{im\theta}, -\infty \leq m \leq \infty
\end{equation}
In Eq.\,(12), $H_m^{(1)}$ is the Hankel function of the first kind of order $m$ and $b_m$ is the corresponding scattering coefficient for an incident wave of the form $J_m(kr)e^{im\theta}$. The field mode with index $m$ has an angular momentum $m\hbar$ along $z$. As expected from the conservation of angular momentum and also following from Eqs.\,(10,11), the scattering coefficients $b_m$ are only non-zero for $m = \pm m_0$. We compute these unknown scattering coefficients using appropriate boundary conditions as done in the previous work \cite{rituraj1}. Now we employ the DtN map based technique to determine the expansion coefficients $c_m$ of the Bloch eigenmode field $\phi_\textbf{q}(\textbf{r})$ in the computed local basis set (Eq.\,(13)). 
\begin{equation}
\phi_\textbf{q}(\textbf{r}) = \sum_{m = -\infty}^{\infty}c_m\phi_{k,m}(\textbf{r})
\end{equation}
This computational procedure is exactly the same as in \cite{rituraj3} and is briefly outlined here. A DtN map is a linear operator which maps the field values at the unit cell boundary to its normal derivative \cite{yuan2006photonic, yuan2007computing, liu2011efficient}. Using the DtN map and applying the Bloch boundary condition for the field (Eq.\,(14a)), one formulates a linear eigenvalue problem, solving which yields the Bloch wavevector $\textbf{q}$ and the Bloch eigenmode coefficients $c_m$. The Bloch condition for excited state amplitudes (Eq.\,(14b)) directly follows from Eqs.\,(11 \& 14a) and thus it suffices to enforce the Bloch condition on the field only.

\begin{subequations}
	\begin{tabularx}{0.9\hsize}{@{}XX}
		\begin{equation} \phi_\textbf{q}(\textbf{r}+\textbf{r}_{uv}) = e^{i\textbf{q}\cdot\textbf{r}_{uv}}\phi_\textbf{q}(\textbf{r}), 
		\end{equation} & \begin{equation}
			e_{\pm uv} = e^{i\textbf{q}\cdot\textbf{r}_{uv}}e_{\pm 00}.
		\end{equation}
	\end{tabularx}		
\end{subequations}

\noindent
Note that the DtN map based technique is different from the usual bandstructure computation where the eigenvalue problem is formulated for a given Bloch vector and yields $\omega^2$ as the eigenvalue. Here, the DtN map is computed for a given SPP frequency $\omega$ (or SPP wavevector $k$) and solving the eigenvalue problem gives the Bloch wavevector $\textbf{q}$ and the corresponding eigenmode. To compute the Berry curvature and Chern number associated with a band, we first use the DtN map based procedure to obtain the Bloch eigenmodes at a discrete set of points in the first Brillouin zone and then use the standard four-point formula for extracting the Berry-Pancharatnam-Zak phase \cite{blanco2020tutorial, marzari1997maximally}.

\section{Results}
We now use the DtN map based formalism described in the previous section to explore several atomic lattices with different topological properties and show some important applications related to non-reciprocal photon transport and unidirectional emission from a single photon source.

\subsection{Topologically trivial lattice $(m_0 = 0)$}

\begin{figure}
\includegraphics[width = 0.53\linewidth, valign=t]{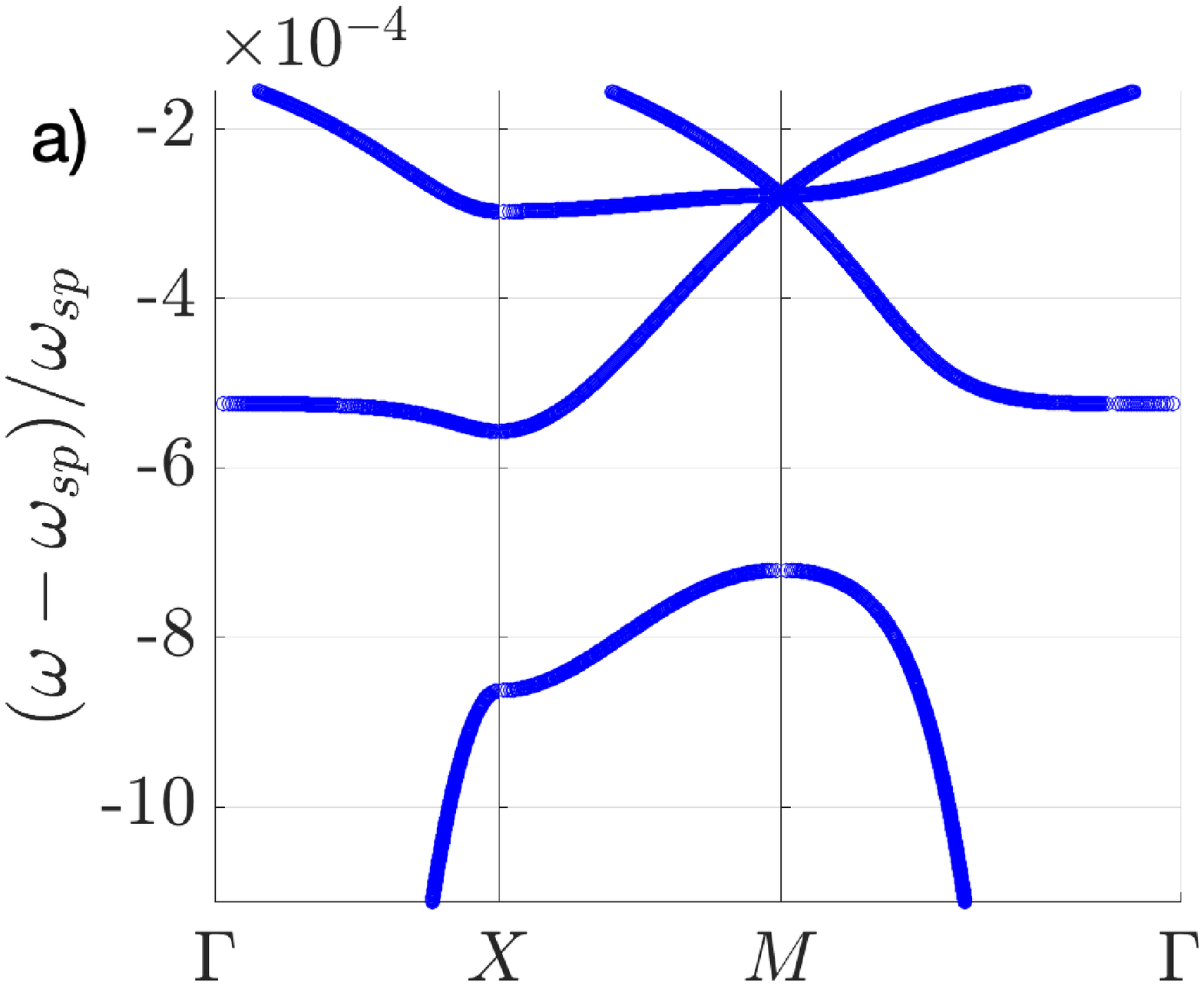}
\includegraphics[width = 0.47\linewidth, valign=t]{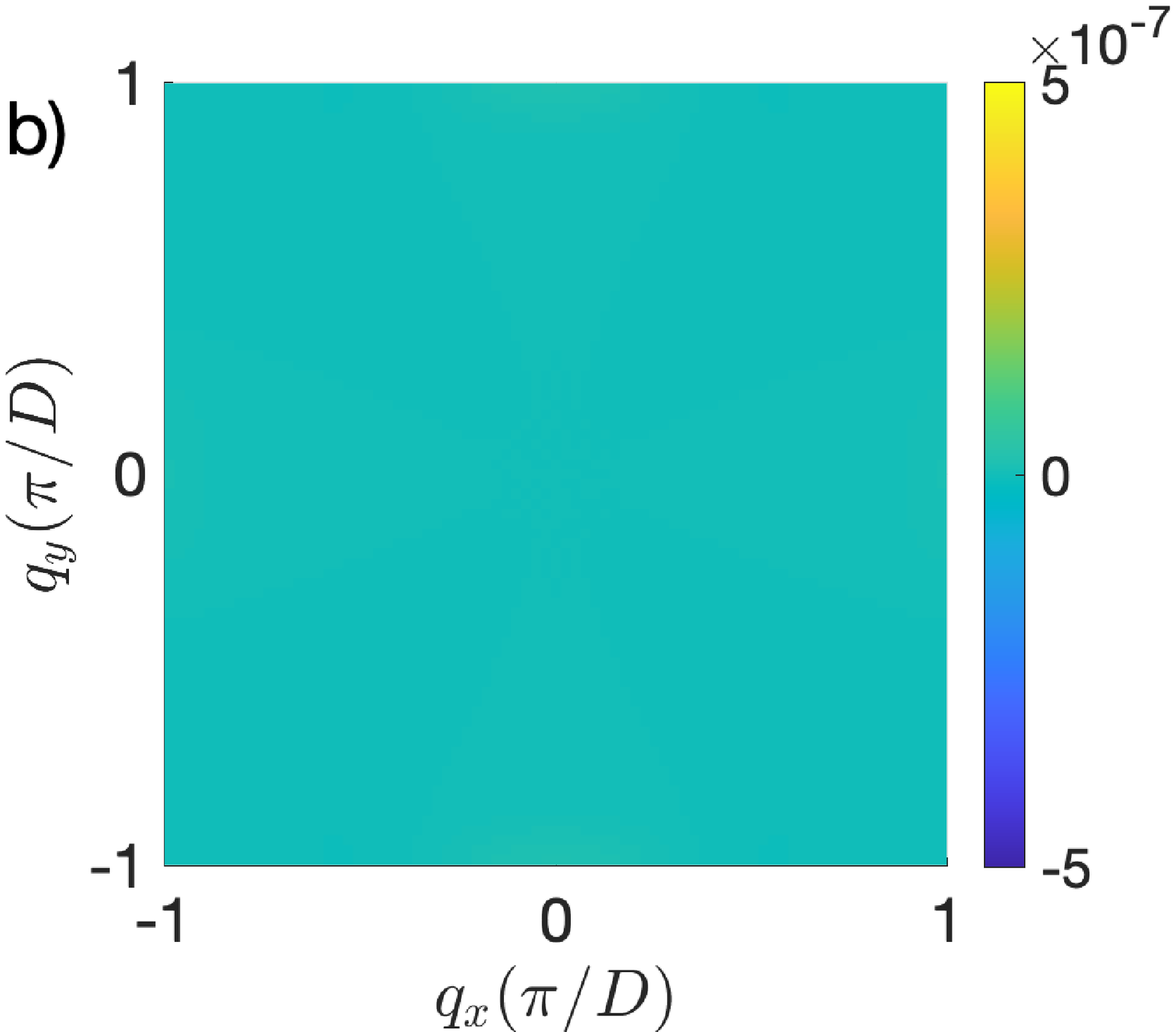}
\caption{(a) Band structure of a square lattice of atoms ($m_0 = 0$) with period $D = 56\,nm$ placed at a distance $h = 10\,nm$ from the surface, (b) Berry flux associated with the lowest band.}
\end{figure}
We start by considering a square lattice of two-level atoms $(m_0 = 0)$ comprised of quantum dots of radius $a = 4\,nm$ and height $L = 1.47\,nm$ placed at a distance of $h = 10\,nm$ from a surface supporting a SPP mode. We set the surface plasmon energy $\hbar\omega_{sp} = 0.5226\,eV$, such that it is resonant with the atomic transition energy $\hbar\Omega = E_{2,0,1} - E_{1,0,1}$. The lattice period is set to $D = 56\,nm$ which sets the Brillouin zone boundary at $\pm(\pi/D = 21.2\omega_{sp}/c)$, where $\omega_{sp}/c$ is the free space wavevector magnitude at the surface plasmon frequency. The SPP wave in the infrared spectrum, with wavelength reduction by a factor of $21.2$ over free space wavelength and low propagation loss, should be easily achievable with conventional noble metals as well as novel 2D materials such as graphene and hBN \cite{jablan2013plasmons, woessner2015highly, fei2011infrared, lee2020pushing, wang2020manipulating}. As we will see later, the SPP propagation length only needs to be several lattice periods to observe the effects emergent from the periodic structure. Fig.\,2a shows the bandstructure of the lattice along the boundary of the irreducible Brillouin zone. The lowest band is isolated by a finite bandgap which arises due to the scattering of the $m = 0$ field mode (Eq.\,(12)). The bandgap is centered around the scattering resonance frequency and its width depends on the scattering linewidth of the atom \cite{rituraj3}. We also see a finite Lamb shift which moves the scattering resonance away from the atomic transition frequency $\Omega$. Fig.\,2b plots the Berry flux for the lowest band, which can be seen to be trivially zero throughout the first Brillouin zone. As a consequence, the Chern number is also $0$. In Section 3.3, we use this lattice to design unidirectional edge states and the lattice parameters chosen here were optimized in order to get a large topological gap.
 
\subsection{Chern insulator $(m_0 = 1)$}

\begin{figure}
\includegraphics[width = 0.53\linewidth, valign=t]{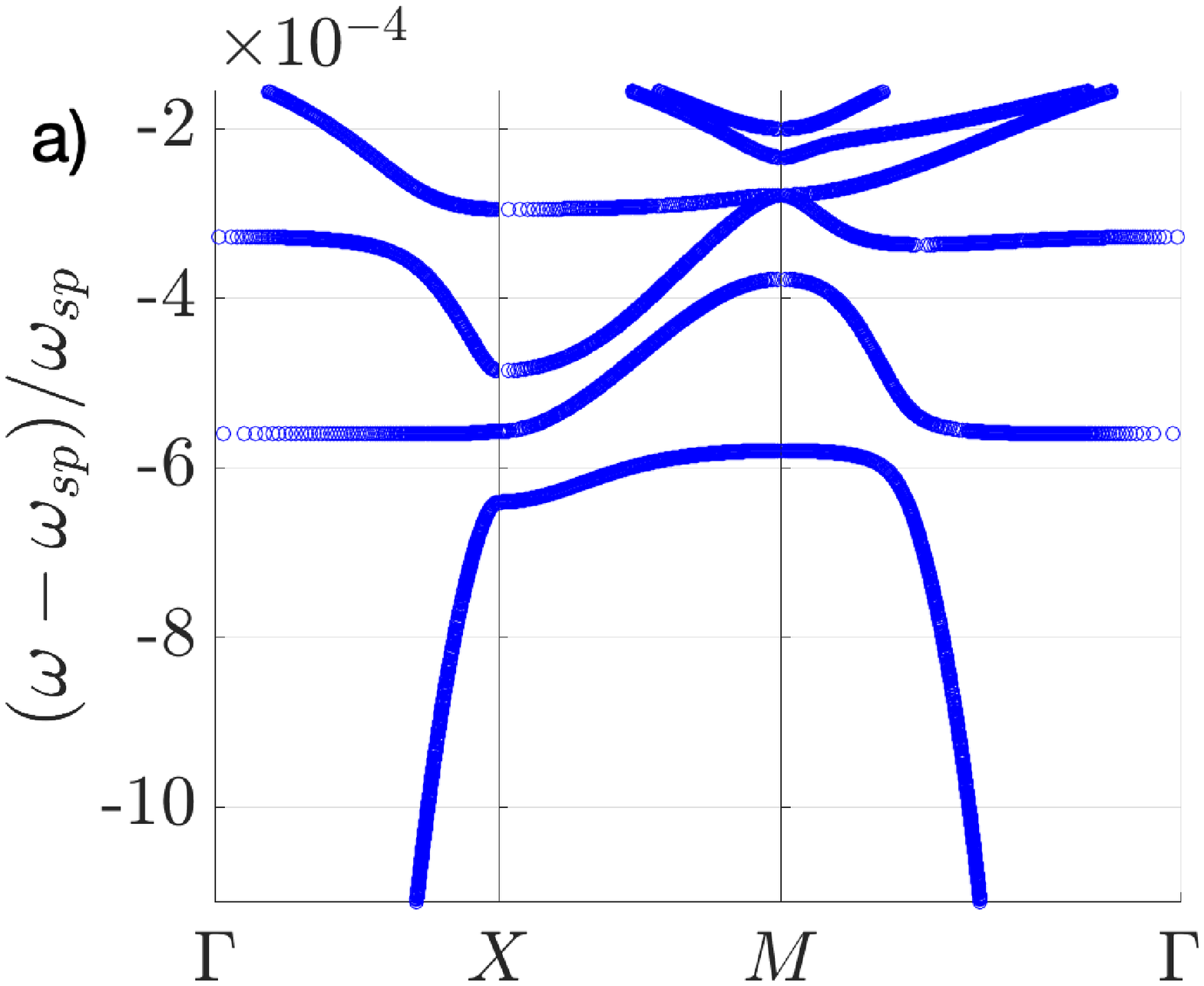}
\includegraphics[width = 0.47\linewidth, valign=t]{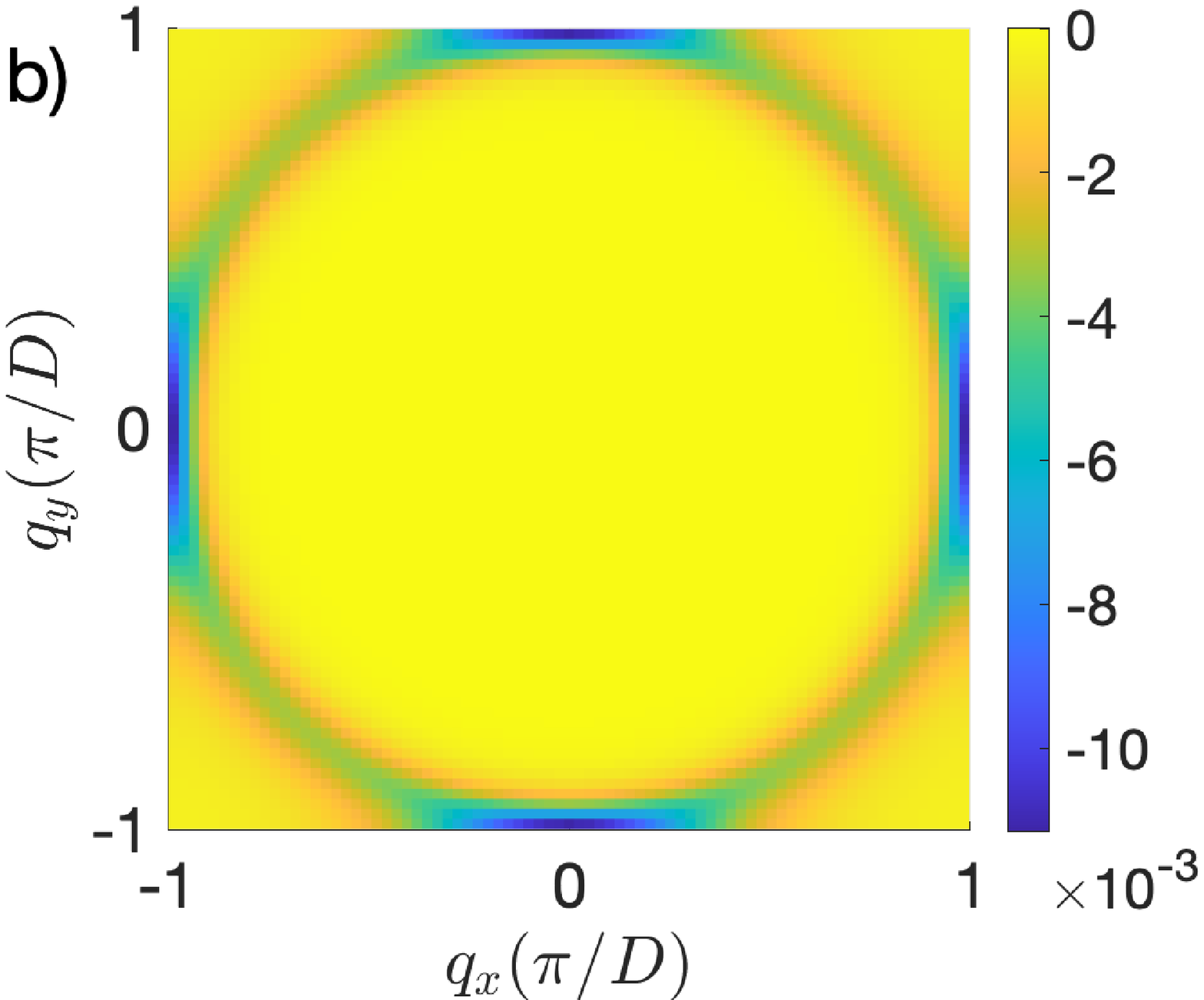}
\caption{(a) Band structure of a square lattice of atoms ($m_0 = 1$) with period $D = 56\,nm$ placed at a distance $h = 10\,nm$ from the surface, (b) Berry flux associated with the lowest band.}
\end{figure}

Now we consider a square lattice of three-level atoms $(m_0 = 1)$ with the same lattice period and surface plasmon frequency as the previous $(m_0 = 0)$ lattice. We choose different quantum dot parameters $a = 4\,nm$, $L = 1.5\,nm$, such that the surface plasmon energy is now resonant with the atomic transition energy $\hbar\Omega_\pm = E_{2,\pm1,1} - E_{1,0,1}$. We apply an external magnetic field of $-4.5\,T\,\hat{\textbf{z}}$ to split the $\ket{+}$ and $\ket{-}$ states by an amount $(\Omega_- -\Omega_+) = 10^{-3}\omega_{sp}$. All these parameters were optimally chosen to obtain a large topological gap which we discuss now. The resulting bandstructure is plotted in Fig.\,3a. Similar to the $(m_0 = 0)$ lattice, the lowest band here is isolated from other bands by a finite bandgap. The bandgap is almost an order of magnitude smaller than the previous lattice because of narrower linewidth associated with the scattering of $m = 1$ field mode as compared to $m = 0$ mode of the previous case. Here, the scattering of $m = -1$ field mode which occurs via transition to $\ket{-}$ state is far detuned from the bandgap and does not affect the lowest band. The scattering linewidth in general decreases for higher $m$ values as the interaction with higher $m$ field modes involves higher order electric and magnetic multipole transitions. As discussed earlier, our model correctly accounts for contributions from electric and magnetic multipole transitions of all orders. 

Fig.\,3b shows the Berry flux for the lowest band, computed by discretizing the first Brillouin zone into a $80\times80$ grid. The Berry flux is concentrated in a thin ring near the Brillouin zone boundary around the resonant frequency for the $m = 1$ field mode scattering. The Chern number turns out to be $-1$. Switching the direction of the external magnetic field does not change the band dispersion (Fig.\,3a) and just flips the sign of the Berry flux and the Chern number.

\subsection{Unidirectional edge states}
\begin{figure}
\includegraphics[width = \linewidth]{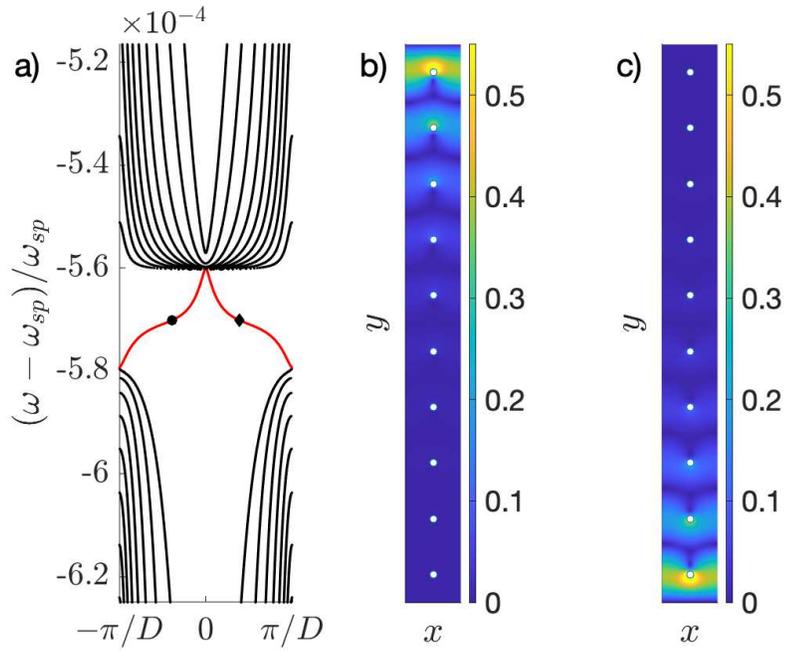}
\caption{(a) Band structure of a 2D lattice of atoms ($m_0 = 1$) with period $D = 56\,nm$ along $x$ and 10 periods along $y$. Field amplitude plots $|\phi_q(\textbf{r})|$ corresponding to the (b) Top, and the (c) Bottom edge modes, at SPP wavevector $k = 20.94k_{sp}$.}
\end{figure}

From the bulk-edge correspondence, we expect that the $m_0 = 1$ lattice, when terminated, can support unidirectional edge states in the topological bulk bandgap. To show this, we consider a 2D atomic lattice with infinite periodic structure along the $x$ direction and 10 lattice periods along the $y$ direction. We assume zero boundary condition at the lattice termination edges $(y = \pm 5D)$ and use the DtN technique to compute the band dispersion and the Bloch eigenmodes. The band dispersion is plotted in Fig.\,4a. As compared to the bulk bandstructure (Fig.\,3a), there are two additional bands inside the bulk band gap which are plotted in red. The two bands correspond to the top and bottom edge modes and have group velocity with opposite signs. The left (right) band corresponds to the top (bottom) edge mode and propagates along $+x$ $(-x)$ direction. Figs.\,4b and 4c show the field amplitude plots $(|\phi_q(\textbf{r})|)$ inside a unit cell for the top and bottom edge modes respectively at SPP wavevector magnitude $k = 20.94k_{sp}$ marked by a circle and a diamond on the band dispersion plot in Fig.\,4a. These unidirectional edge modes persist even for a smaller lattice size. This is illustrated in Fig.\,5 which shows the top and bottom edge states at the same SPP frequency but in a smaller lattice with only 3 lattice periods along $y$.

\begin{figure}
\centering
\includegraphics[width = 0.7\linewidth]{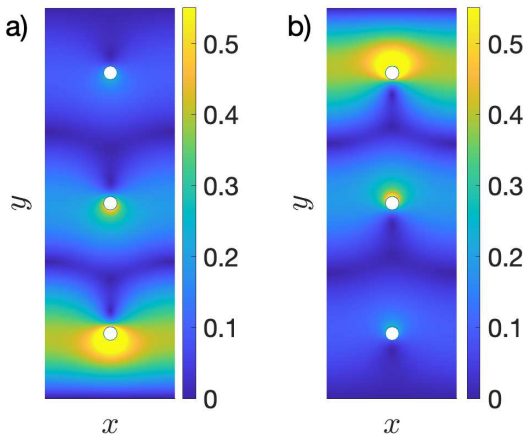}
\caption{Field amplitude plots $|\phi_q(\textbf{r})|$ corresponding to the (a) Top, and the (b) Bottom edge modes of a 2D lattice of atoms ($m_0 = 1$) with period $D = 56\,nm$ along $x$ and 3 periods along $y$, at SPP wavevector $k = 20.94k_{sp}$.}
\end{figure}

\begin{figure}[ht]
\includegraphics[width = \linewidth]{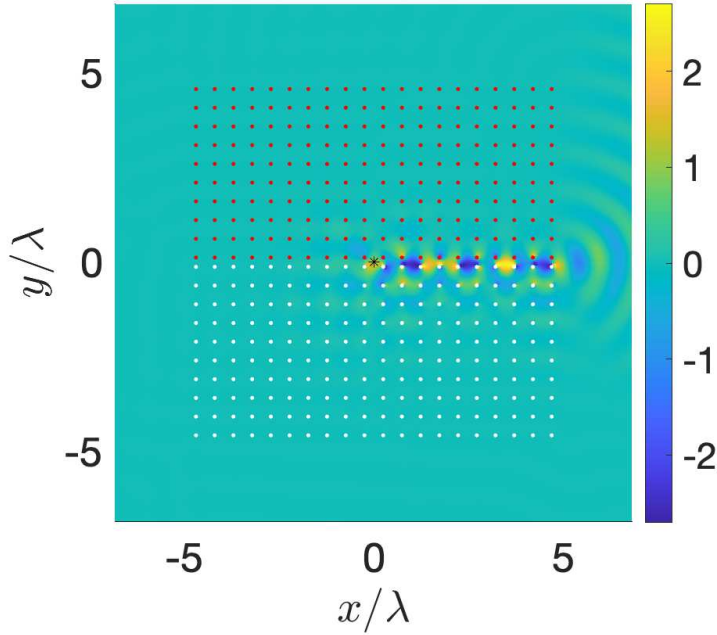}
\caption{Field plot $Re(\phi(\textbf{r}))$ for single photon emission from a point source placed at origin at the interface between $20\times10$ unit cells of $m_0 = 0$ lattice $(y>0)$ and $20\times10$ unit cells of $m_0 = 1$ lattice $(y<0)$. The emitted photon has a frequency corresponding to the SPP wavevector magnitude $20.94k_{sp}$.}
\end{figure}
One can harness these topologically robust edge states to achieve unidirectional emission as well as non-reciprocal transmission of single SPP photon for frequencies inside the topological gap. To this end, we consider an interface between the topologically distinct $m_0 = 0$ and $m_0 = 1$ lattices, described in the previous sections. As shown in Fig.\,6, we consider a finite structure with $20\times10$ unit cells of $m_0 = 0$ lattice placed at $y > 0$ and $20\times10$ unit cells of $m_0 = 1$ lattice placed at $y<0$. The atomic positions in the two lattices are marked by red and white points respectively. By design, the bandgap of the $m_0 = 1$ lattice with Chern number $-1$ falls within the bandgap of the topologically trivial $m_0 = 0$ lattice and thus the interface supports a single right propagating edge mode. Since the DtN technique is no longer applicable for a finite lattice, we compute the field profile using multiple scattering formalism proposed in our previous work \cite{rituraj2} which is quite general and works for an arbitrary number of atoms/unit cells. It involves first computing the scattering from an isolated atom/unit cell and then expressing the scattered field from the finite lattice in terms of the scattered field modes from each unit cell. Fig.\,6 shows the field plot $(Re(\phi(\textbf{r})))$, where the edge mode is excited by a point source placed at origin (marked by star) and oscillating at frequency corresponding to the SPP wavevector magnitude $k = 20.94k_{sp}$. It clearly shows that the oscillating point source emits a SPP photon into the edge mode.

\begin{figure}
\includegraphics[width = \linewidth]{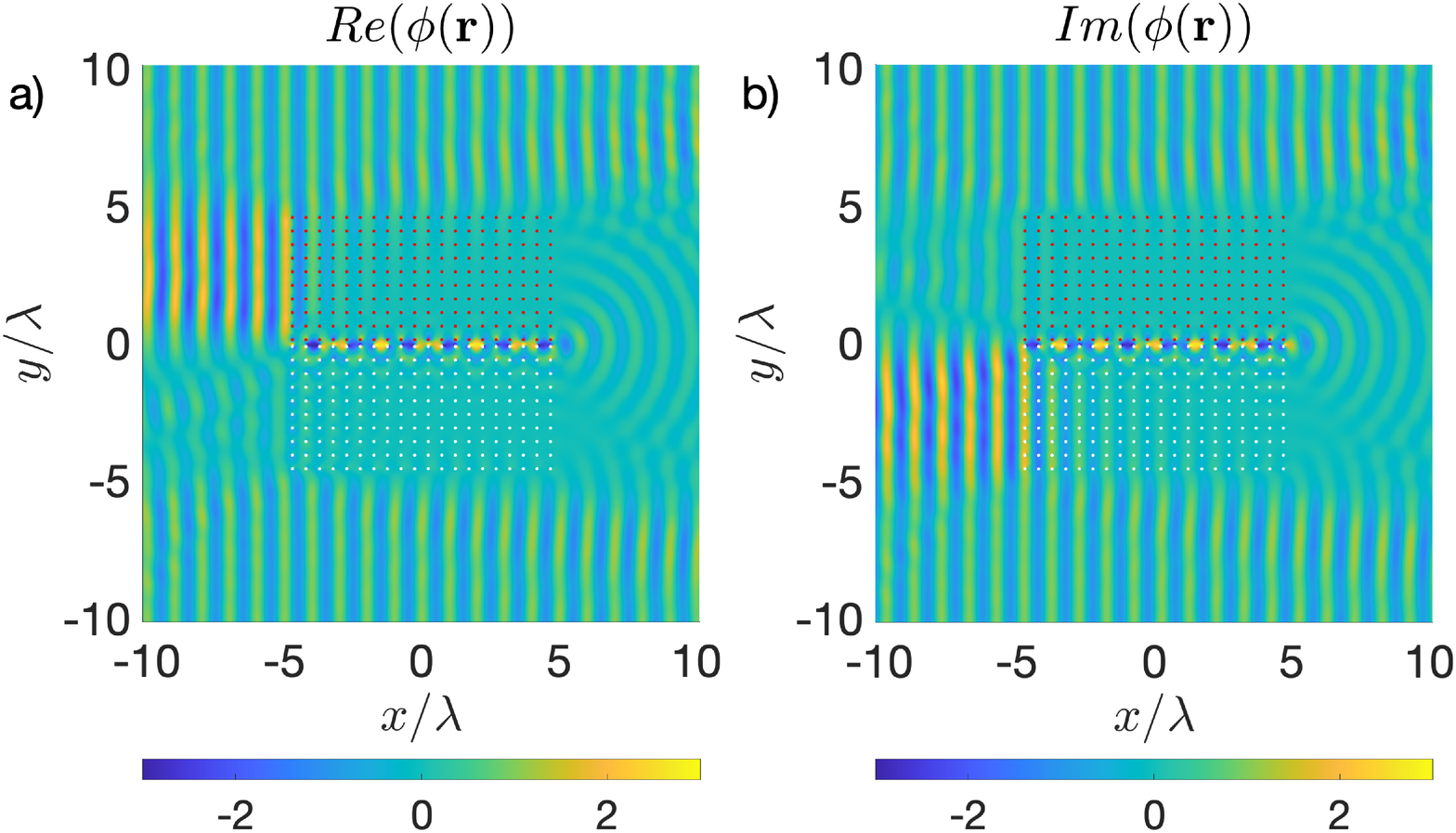}
\includegraphics[width = \linewidth]{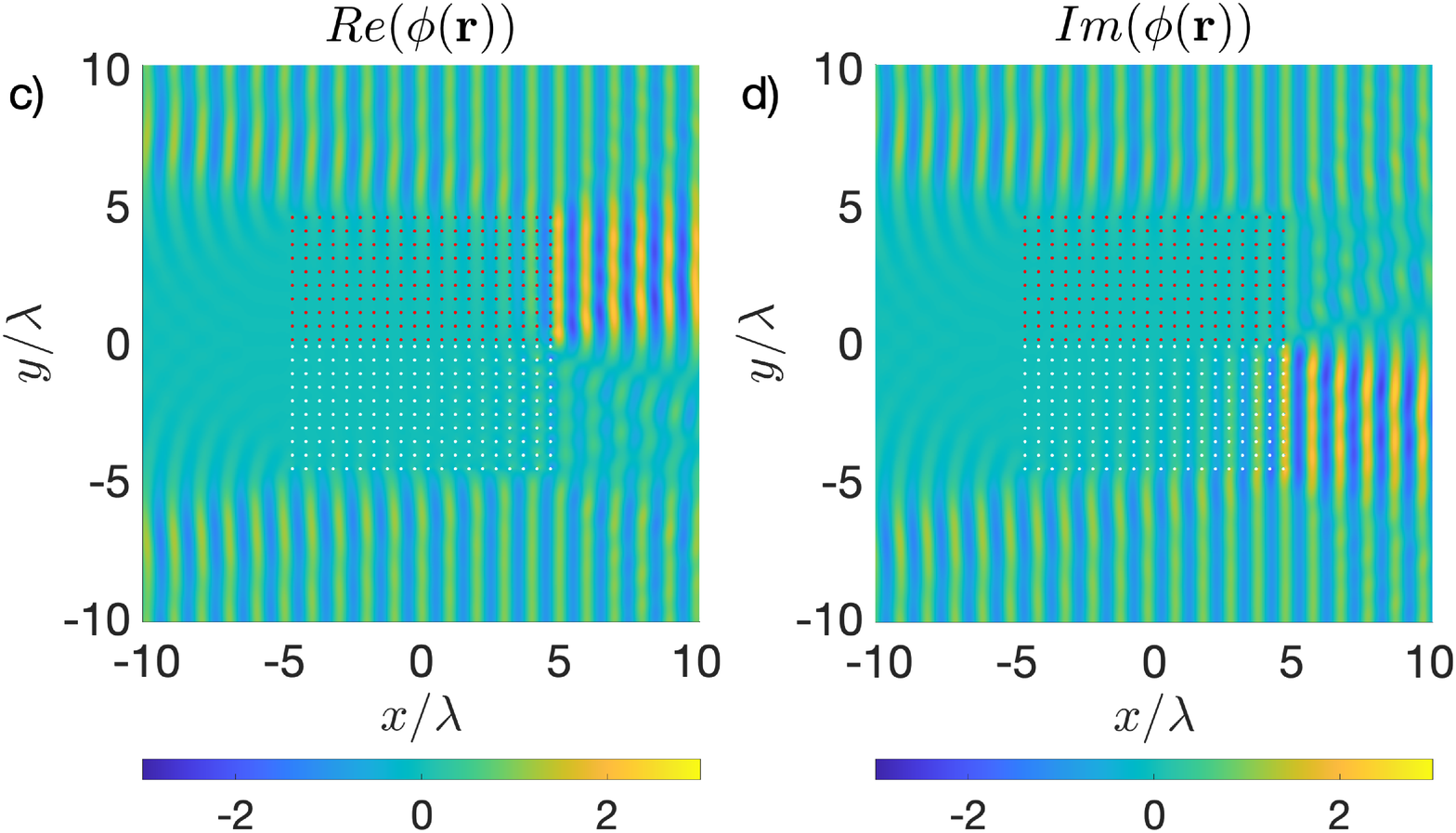}
\caption{Field plots $Re(\phi(\textbf{r}))\,\&\, Im(\phi(\textbf{r}))$ for a SPP plane wave with wavevectors (a, b) $\textbf{k} = 20.94k_{sp}\hat{\textbf{x}}$ and (c, d) $\textbf{k} = -20.94k_{sp}\hat{\textbf{x}}$, incident upon a finite structure with $20\times10$ unit cells of $m_0 = 0$ lattice $(y>0)$ and $20\times10$ unit cells of $m_0 = 1$ lattice $(y<0)$.}
\end{figure}

Using this unidirectional edge state, one can also realize non-reciprocal photon transmission as shown in Fig.\,7. Figs.\,7(a,b) plot the real and imaginary components of the total field profile when a single photon in SPP plane wave mode is incident from the left at the same frequency as the previous case $(\textbf{k}_{inc} = 20.94k_{sp}\hat{\textbf{x}})$. The incident photon couples to the edge mode at the interface and gets transmitted. Due to the absence of bulk modes at this frequency in both top and bottom lattices, there is no transmission through the lattice away from the interface. Figs.\,7(c,d) plot the real and imaginary components of the total field when the photon is incident from the right, in the $-x$ direction at the same frequency. Due to the unidirectional nature of the edge mode, in this case there is no transmission through the interface. Also, the incident plane wave can be seen to gain different phases upon reflection from the top and bottom lattices. This happens because the atoms in the top lattice $(y > 0)$ scatter $m = 0$ field modes whereas the atoms in the bottom lattice $(y < 0)$ scatter $m = 1$ field modes (Eq.\,(12)). The scattered field is given by the Hankel function of order $0$ and $1$ respectively which have a $\pi/2$ phase difference in the far field.  

\subsection{Larger Chern number $(m_0 = 2)$}
\begin{figure}
\includegraphics[width = 0.5\linewidth]{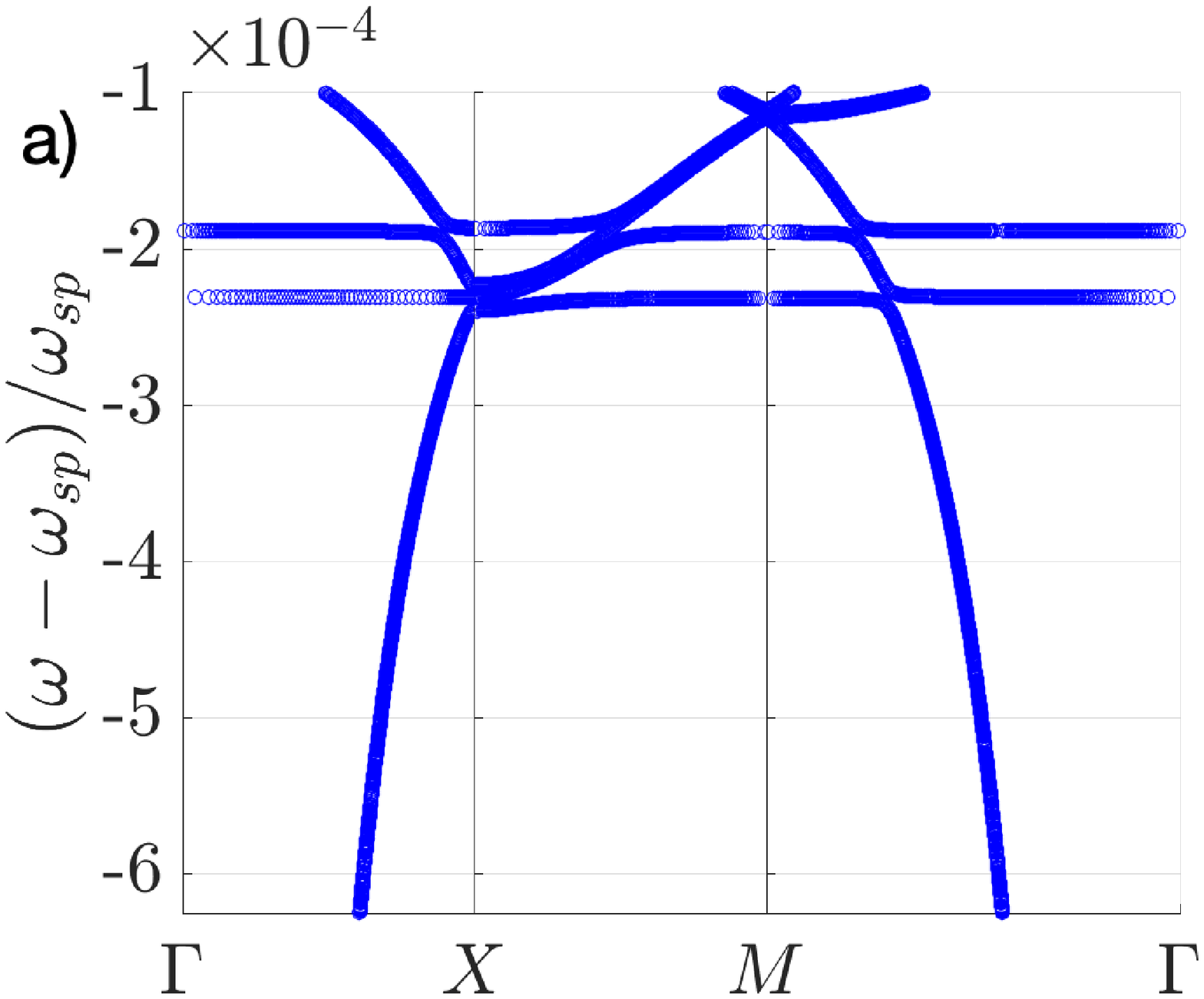}
\includegraphics[width = 0.5\linewidth]{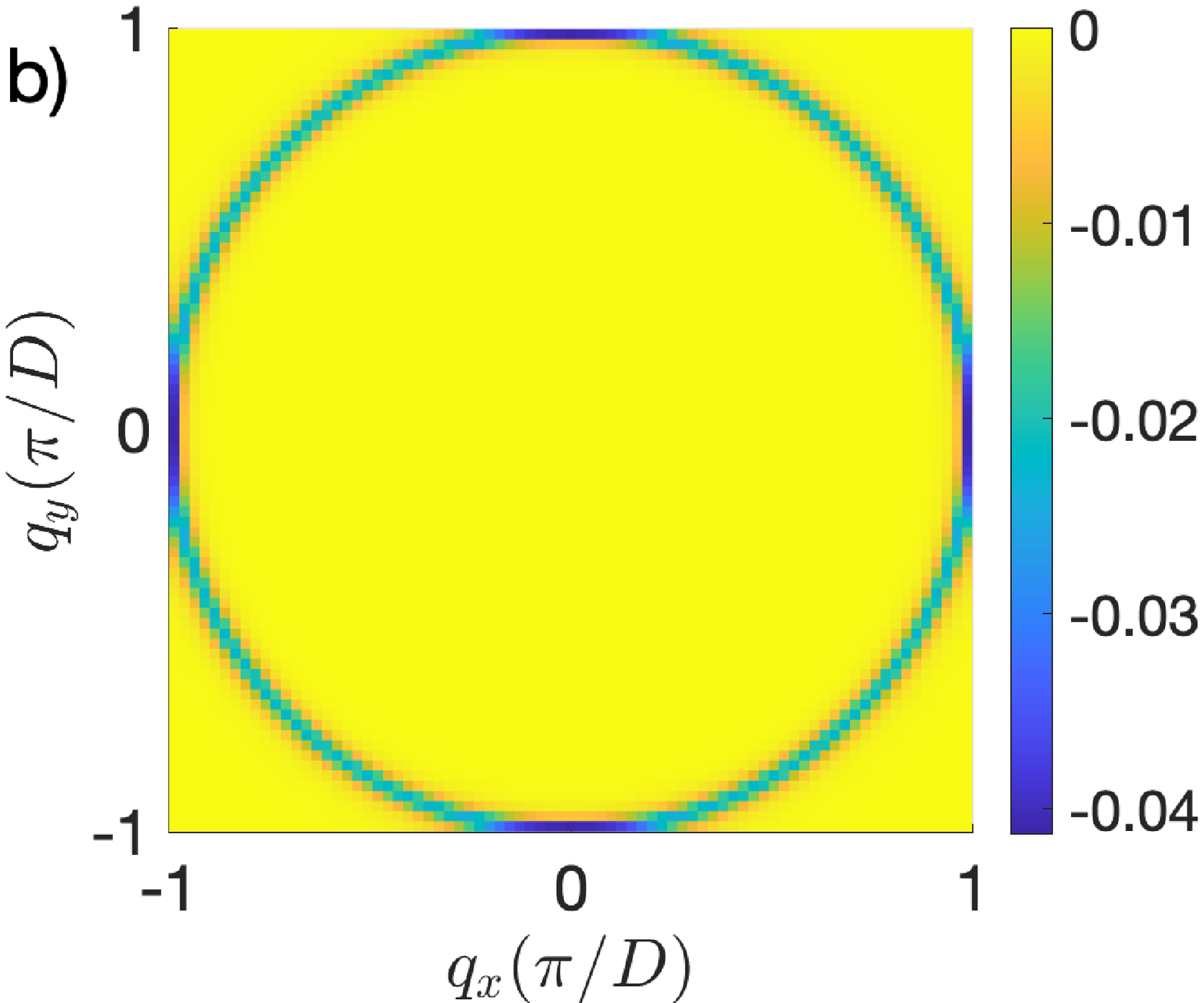}
\caption{(a) Band structure of a square lattice of atoms ($m_0 = 2$) with period $D = 36\,nm$ placed at a distance $h = 5\,nm$ from the surface, (b) Berry flux associated with the lowest band.}
\end{figure}
It is possible to achieve higher Chern numbers by having atomic transitions to excited states with higher angular momentum quantum number $m_0$. In the general case, one can obtain a Chern number of $\pm m_0$ depending upon the direction of the applied external magnetic field. Here, we illustrate this for the case of a square lattice of three level atoms with $m_0 = 2$, i.e.,\,excited states $\ket{\pm}\equiv (2, \pm2, 1)$, and point out some of the challenges associated with realizing topological gaps with higher Chern numbers. For comparison, the surface plasmon frequency $(\omega_{sp} = 0.52\,eV/\hbar)$ and the external static magnetic field $(-4.5\,T\,\hat{\textbf{z}})$ are kept similar to the previous case of $m_0 = 1$ lattice. As discussed earlier, the linewidth associated with the scattering of $m = 2$ field mode is much smaller than that for $m = 1$ mode. To enhance this linewidth and obtain a reasonable topological gap, we place the atoms closer to the surface $(h = 5\,nm)$ and choose the quantum dot parameters $(a = 6\,nm, L = 1.5\,nm)$ such that the scattering resonance frequency lies closer to the surface plasmon frequency. For these parameters, the scattering resonance occurs at frequency corresponding to the SPP wavevector magnitude $k \approx 32.9k_{sp}$. Note that this SPP confinement factor of $32.9$ is larger than the previous atomic lattices and thus puts more stringent constraints on the system since the SPP propagation loss typically increases with higher confinement. Furthermore, opening a complete band gap requires the first Brillouin zone edge $(\pi/D)$ to be larger than the resonant SPP wavevector $32.9k_{sp}$. This requires a smaller lattice period and here we choose $D = 36\,nm\,(\approx \pi/(32.9k_{sp}))$. Even with these optimized parameters, the scattering linewidth and the bandgap is more than an order of magnitude smaller than that for the $m_0 = 1$ lattice. The bandstructure is plotted in Fig.\,8a. The lowest band is isolated by a much smaller bandgap. Similar to the case of $m_0 = 1$ lattice, transition to the $\ket{-}$ excited state is highly detuned due to the Zeeman splitting and the width of the bandgap only depends on the linewidth associated with the scattering of $m=2$ field mode which involves transition to $\ket{+}$ excited state. Fig.\,8b plots the Berry flux associated with the lowest energy band and is qualitatively similar to the case of $m_0 = 1$ lattice. The Berry flux is concentrated in a thin ring near the Brillouin zone boundary around the resonant frequency for the $m = 2$ field mode scattering. The Chern number is $-2$ in this case.

\section{Conclusion}
We have studied various 2D atomic lattices interacting with a single photon in the surface plasmon polariton mode, and showed the possibility of realizing bands with different topologies. We also demonstrated unidirectional single photon emission and single photon transmission via topologically robust edge states at the interface between two lattices with different Chern numbers. Finally, we pointed out how one could realize topological gaps with arbitrary Chern number by manipulating the internal degree of freedom (angular momentum quantum number $m_0$) of the atoms, and also discussed some of the challenges, namely smaller lattice periods and interaction with shorter wavelength plasmons, in designing systems with higher Chern numbers.

\section*{Acknowledgement}
This work is supported by a Vannevar Bush Faculty Fellowship from the U.\ S. Department of Defense (Grant No. N00014-17-1-3030). Rituraj acknowledges the support from a Stanford Graduate Fellowship.

\bibliography{ref}
\bibliographystyle{unsrt}

\end{document}